# Impacts of suppressing guide on information spreading

Jinghong Xu[1], Lin Zhang[2], Baojun Ma[3], Ye Wu[2,*]
[1] School of Digital Media and Design Arts, Beijing University of Posts and Telecommunications, Beijing 100876, China
[2] School of Science, Beijing University of Posts and Telecommunications, Beijing 100876, China
[3] School of Economics and Management, Beijing University of Posts and Telecommunications, Beijing 100876, China
[*] Corresponding author, email: wuye@bupt.edu.cn

## Abstract
It is quite common that guides are introduced to suppress the information spreading in modern society for different purposes. In this paper, an agent-based model is established to quantitatively analyze the impacts of suppressing guides on information spreading. We find that the spreading threshold depends on the attractiveness of the information and the topology of the social network with no suppressing guides at all. Usually, one would expect that the existence of suppressing guides in the spreading procedure may result in less diffusion of information within the overall network. However, we find that sometimes the opposite is true: the manipulating nodes of suppressing guides may lead to more extensive information spreading when there are audiences with the reversal mind. These results can provide valuable theoretical references to public opinion guidance on various information, e.g., rumor or news spreading.

## Keywords
Information spreading, reversal psychology, public opinion guidance, complex networks

## Introduction
Exploring the dynamics of information spreading and disease propagation is an important topic, which has attracted increasing attentions in recent years [1-4]. The spreading of information may be influenced by social reinforcement and public opinion guidance [4-10]. Social reinforcement is defined as the situation in which an individual, before adopting an opinion, requires multiple prompts from his or her neighbors [11]. Introducing guide into the spreading system is a common phenomenon in our daily life [12], especially during the period of emergencies [13, 14]. Guides are very common and active in viral marketing while viral messages are playing an important role in influencing and shifting public opinions about corporate reputations, brands, and products as well as political parties and public figures, etc. [15]. However, the results of opinion-guiding may not always be the same as desired. Contrary to prompting the spreading of the information for certain, the suppressing effect of the guide may sometimes lead to the reversal of the audience attitude [16, 17], and eventually decreases the probability of extensive information spreading. While there are abundant qualitative analysis of the effects of opinion-guiding, mainly based on the principles of mass media, psychology and other social sciences [10, 11, 17, 18], quantitatively analyzing the suppressing effect and the reversal effect still remains a crucial and urgent topic in the field of information spreading.

In this paper, we focus on two factors, *the suppressing guides* whose effect is to decrease the probability of their network neighbors adopting the information as well as *the reversal mind of audiences* whose effect is to provoke information spreading when there is suppressing guiding nodes. We found that if there is no guide, the breaking point of information depends mainly on the network topology and the attractiveness of the information. As different rates of the suppressing guides appear, stimulating, ineffective or inhibitory impacts may occur under different conditions. Finally, the effect of "reverse psychology" and stimulation of information spreading caused by suppressing guide are investigated and revealed quantitatively. This paper is organized into four parts. The first part is the introduction. The second part presents our model and the third part quantitatively analyzes the information spreading procedure. The fourth part includes conclusions and discussions.

## Model

Model studies usually aim to reproduce some empirical observations to uncover the main mechanisms of the underlying processes. Dynamic process of complex systems can be considered as one taking place on a network formed by pairwise interactions between the constituents of the system [19, 20], and the information spreading takes place within the network.

Our model consists of a random network of $N$ nodes, representing $N$ participants in the information spreading system. At each discrete time step $t$, each node $k$ may be in one of the two states $S_k(t)=0$ or $S_k(t)=1$, representing unknown/non-acceptance or adoption of certain information, respectively. When a node $k$ is in the adoption state $S_k(t)=1$, node $m$ receives an input of strength $A_{km}$ from $k$. Each node $k$ is either an ordinary person or a suppressing guide, corresponding to $A_{km}>0$ or $A_{km}<0$ for all $m$. Negative strength means the prevention of the information spreading by suppressing guides. If there is no connection between node $k$ and node $m$, then $A_{km}=0$. At time $t+1$, the state of node $n$ switches as a Markovian process with the following transition rule:

$$S_n(t+1) = 1, \text{ with probability } \sigma\left(\sum_{m=1}^{N} A_{mn} S_m(t)\right),$$

and $S_n(t+1)=0$, otherwise, where the transfer function is piecewise linear defined as

$$\sigma(x) = \begin{cases} 0, & x \leq 0, \\ x, & 0 < x < 1, \\ 1, & x \geq 1. \end{cases}$$

Intuitively, in our model, the "adoption" of an agent is the comprehensive effect of all its neighbors. Analogously, an "adoption" agent changing the state to "unknown or unconvinced" also depends on the effect of all their neighbors. In fact, no matter what states (0 or 1) the agents are in, the probability of choosing state 1 depends on the states of all their neighbors, and so is the probability of choosing state 0. For example, in the social media such as micro-blog or WeChat, one's adopting and forwarding a certain message is merely the effect of his/her "neighbors" list in the circle of friends. According to our model, when there is no adoption node, or when there are suppressing guiding nodes in the network, a node would never adopt the rumor. Moreover, larger ratio of the suppressing guiding nodes around would lead to lower probability of a node's adopting the information since the $A_{km}$ is negative if node $k$ is a suppressing guide. On the other hand, if there is no suppressing guiding node in the network, more adoption nodes around will lead to larger probability of a node's switching from unknown/non-acceptance to adoption [21].

We consider the dynamics described above on a directed random network. The connecting probability of each pair of nodes is $p$. Each nonzero connection strength $A_{km}$ is independently drawn from a uniform distribution on $[0, 2r]$, with mean strength $r$. The larger the value of $r$ is, the more popular or attractive the information shows. Next, a fraction of the nodes $\alpha$ is designated as suppressing guides and each row of the matrix $A$ that corresponds to the outgoing connections of a suppressing guiding node is multiplied by -1.

In this work, we focus on the average adoption nodes of the network, defined as

$$S(t) = \frac{1}{N}\sum_{n=1}^{N} S_n(t), \qquad (1)$$

which is the fraction of nodes that is of the adoption state at time t. According to Eq. (1), if the entire network does not know the information, $S=0$, it will remain unknown indefinitely. In the following, we will investigate the effect of parameters $r$ and $\alpha$, namely, the effect of the attractiveness of information and the ratio of suppressing guides.

## Results

In the following, we will illustrate our results in the cases of $\alpha=0$, and $\alpha>0$, respectively, corresponding to no suppressing guides at all and with $\alpha N$ suppressing guides. In the case of no suppressing guides, the critical value of connection strength $r$ would be investigated. In the $\alpha N$ suppressing guides case, we compare the situations of no response and the reversal response, aiming at investigating the psychological effect of suppressing guides.

## Case 1: $\alpha=0$.

In the ordinary-people-only case, i.e., α=0, we investigate the effect of average connection strength r during information spreading procedure. As explained above, the connection strength can be regarded as the popularity or attractiveness of the certain information. The unattractive information may die out in a short period of time, while the information would spread widely if it is fascinating enough instead. By considering the above facts, we illustrate the results of rumor spreading for different values of $r$ in Fig. 1. The number of known nodes $N_i$ evolving with time is in consideration. As shown in Fig. 1(a), when $r=0.010$ and $0.012$, the spreading of rumor dies out after a short time. Moreover, smaller $r$ corresponds to shorter spreading time. However, in Fig. 1(b), when $r=0.014$ and $0.016$, the spreading of rumor breaks out to the entire network---almost everyone knows the rumor eventually. Furthermore, larger $r$ corresponds to faster spreading speed. However, t is valuable and crucial to explore the critical value of $r$ that divides the spreading into the two opposite directions.

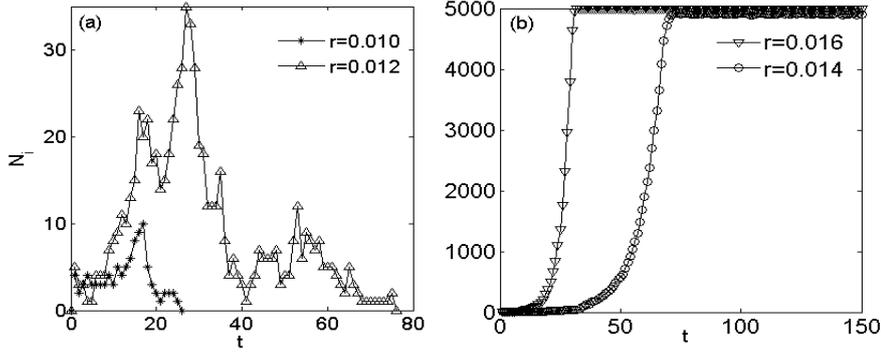

Figure 1. The time series of the number of known nodes evolves with time for different $r$. (a) $r=0.010$ and $0.012$, the spreading dies out eventually. (b) $r=0.014$ and $0.016$, the spreading breaks out.

In the following, the breaking out of any information is identified as it spreads to more than a half nodes of the entire network, i.e., $S(t)>0.5$ for sufficient large $t$. We simulate the spreading process 1000 times for each $r$ and count the ratio of $S(t)>0.5$. Apparently, the ratio changes with the strength parameter $r$. This relationship is illustrated in Fig. 2. It shows that as the increase of the connection strength $r$, the ratio of adoption $R$ stays near zero in the beginning, implying that the information does not spread out. However, when $r$ is greater than a certain value, the ratio $R$ increases sharply as $r$ increases. After the sharp increment, the ratio grows towards 1 stably. The critical value of $r$, say $r_c$, is of great importance in the spreading procedure. Smaller $r_c$ indicates that the information spreads out easily. It is proved in theory [22] that $r_c$ depends on the topology of the network, which is the reciprocal of the mean degree, i.e.,

$$r_c = \frac{1}{\langle k \rangle}.$$

In our simulation, $<k>=80$, thus, the critical value is $r_c=0.0125$ as illustrated in Fig. 2. On the Internet, social networks always possess large average degree, leading to the small critical value of rumor breaking out. Therefore, the prevention of rumor in modern society is thorny to handle.

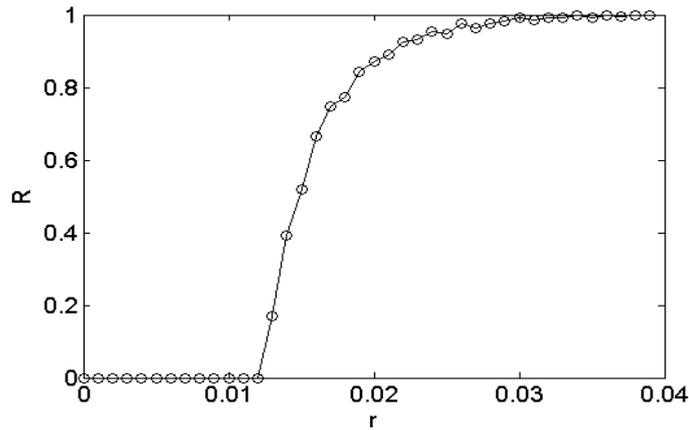

Figure 2. The ratio $R$ that the information spreads over 50% of the entire network changes with the average connection strength $r$. The probability is calculated over 1000 times of simulations.

### Case 2: $\alpha > 0$, no response.

In the following, we will focus on the effect of the ratio of suppressing guides on information spreading, i.e., $\alpha > 0$. Generally speaking, more suppressing guides will lead to the prevention of information spreading when there are no irregular responses. Accordingly, the quantitative analysis based on our model has been proposed as below, which matches common sense well.

In order to investigate the effect of suppression, we need to ensure that the information can spread out within the network. Therefore, the connection strength parameter is fixed as $r=0.02>r_c$. The information spreading procedures with different values of $\alpha$ are demonstrated in Fig. 3. It is shown that different $\alpha$ would lead the information spreading to two completely opposite directions---breaking out or dying out. A natural question is that whether there is any specific critical value of $\alpha$ to determine the two directions. Moreover, how to calculate the specific critical value $\alpha_c$ which ensures the prevention of information spreading?

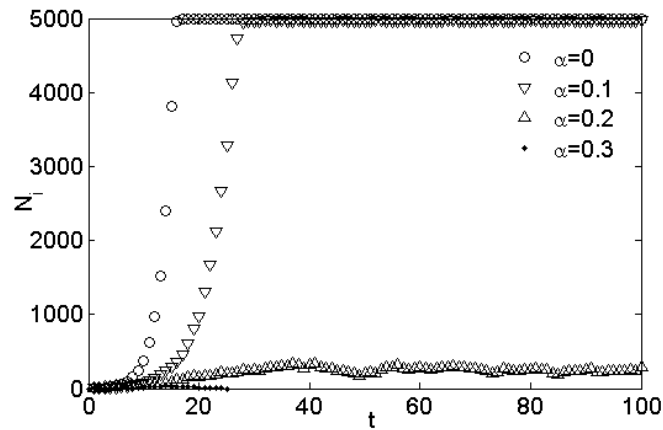

Figure 3. The time series of the number of adoption nodes evolves with time for different $\alpha$.

Next, in Fig. 4, we will present the relationship between the ratio of information breaking out and that of suppressing guides within the entire network. The simulation time is also 1000, and the ratio of $S(t)>0.5$ are calculated for sufficient large $t$. Smaller rate of suppressing guides cannot prevent the outbreak of the information spreading within the entire network. When the ratio is greater than a certain critical value $\alpha_c$, the information spreading can be prevented and none of the 1000 times simulation can spread to more than 50% nodes of the entire network.

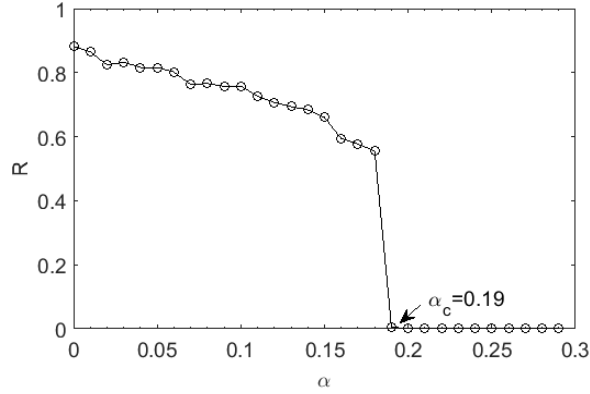

Figure 4. The ratio that the information spreads over 50% of the entire network changes with the suppressing guide ratio $\alpha$ for fixed $r=0.02$. The probability is calculated over 1000 times of simulations.

In order to obtain the critical value $\alpha_c$, we use the accurate approximation of the relationship among the attractiveness $r$, average degree $<k>$, and the suppressing rate $\alpha$ [22,23], which is shown below:

$$r = \frac{1}{\langle k \rangle (1-2\alpha)}. \quad (2)$$

We confirm this relationship by our model in Fig. 5, in which the scatter plot of $1-2\alpha$ versus $r<k>$, together with a fitting curve $y=1/x$ are drawn. The scatter dots are well fitted by the fitting curve, which ensures Eq. (2). Thus the critical value can be expressed as

$$\alpha_c = \frac{1}{2}\left(1 - \frac{1}{r\langle k \rangle}\right).$$

In our simulation, for the fixed $r=0.02$ and $<k>=80$, the critical value is $\alpha_c = 0.19$, which is illustrated above in Fig. 4.

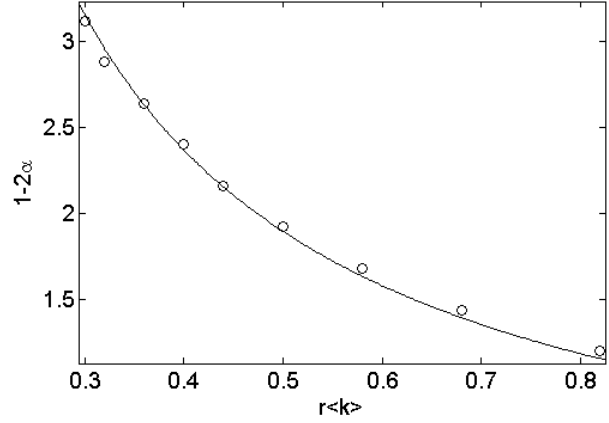

Figure 5. The scatter plot of $1-2\alpha$ versus $r\langle k\rangle$, together with an inverse function as the fitting curve.

The critical value of suppressing guides rate depends both on the attractiveness of the information and the topology of the network. More attractive information and larger average of the network need larger $\alpha_c$ to prevent the outbreak of information spreading.

### Case 3: $\alpha$>0, reversal response.

In this case, we will explore the reversal mind effect of ordinary people and prove that the appearance of suppressing guide may stimulate the information spreading. If the rate of suppressing guides increases, the connection strength may increase as well, implying the reversal of people towards the suppression. If we assume that the connection strength and the suppressing guide ratio possess the following growth function:

$$r = \frac{0.02}{1+200\exp(-50\alpha)}. \tag{3}$$

In Eq. (3), the connection strength $r$ is increasing monotonically with the ratio of guided nodes $\alpha$. We recalculate the high ratio that the information spreads to over 50% nodes of the entire network changes with the suppressing guide rate $\alpha$. As shown in Fig. 6, small rate of suppressing guides leads to small connection strength, and the information dies out since $r$ is lower than the critical value at this stage. After a first critical value, the information breaks out since more suppressing guides lead to reversal mind of people, which results in stronger connection strength. As $\alpha$ increases, the "more than a half spreading" ratio increases, then decreases in this outbreak region. After the second critical value, the information dies out again since the effect of suppressing guides is strong enough to prevent the information spreading although the connection strength is strong. The outbreak phenomenon can be understood as a result of the competition between the increase of the connection strength and the decrease of the number of normal nodes: small connection strength cannot make the rumor spread out, and small number of normal nodes also cannot make the attractive rumor spread out either. Suitable attractiveness may not lead to strong suppressing effect, and therefore the information may break out, which should be paid much attention to.

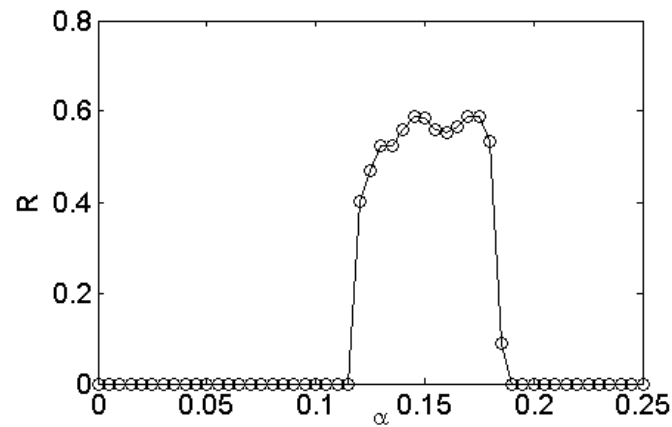

Figure 6. The ratio that the information spreads to over 50% nodes of the entire network changes with the guide ratio $\alpha$ and the connection strength parameter $r$ given as Eq. (3). The ratio is calculated over 1000 times of simulations.

In summary, by considering the reversal effect of suppressing guides, we find that the pattern of dying out-breaking out-dying out in the information spreading under different suppressing guides rate $\alpha$, which may give us theoretical references on the effective control of the outbreak of the information spreading.

## Discussion and conclusions

In this paper, we focus on the information spreading procedure with the effect of suppressing guides. A simple Markovian model is illustrated to describe the information spreading procedure. Three cases of information spreading under different conditions are investigated: no suppressing guides at all, with suppressing guides while the ordinary agents possess no reversal response to the suppression, and with suppressing guides but the ordinary agents possess reversal response to the suppression. Our models and experiments show the following results: (1) when there are no suppressing guides within the entire network at all, the outbreak of information spreading only depends on the topology of the network and the attractiveness of the very information; (2) when there are suppressing guides within the entire network and there is no reversal response of the ordinary agents towards the suppression, the information spreading can be suppressed by the suppressing guides of proper rate; (3) when there are suppressing guides within the entire network, as well as the ordinary agents possess reversal response towards the suppression, a doomed die-out information can be stimulated to break out. Moreover, with the increasing of the rate of suppressing guides, the information spreading can be suppressed eventually. Our theoretical model and the simulation results match common intuitions well, which shed light on the effective control of information spreading. In further works, it is valuable to investigate how to successfully introduce the suppressing guides, for example, the most suitable occasion or time to introduce the suppressing guides, and the most effective suppressing guide to control the information spreading procedure, etc.

## Acknowledgements
This work was jointly supported by Beijing Social Science Fund (Grant No. 15ZHB006) and the National Natural Science Foundation of China (Grant Nos. 71231002, 71402007, and 11426043).